\documentstyle{menu95}
\newcommand{\tcaption}[1]{                      %CENTRALISE TABLE CAPTION
        \addtocounter{table}{1}
         {{\tenrm\offinterlineskip Table~\thetable . #1} }\hfil\break }
\begin{document}
\title{
``STRANGE'' NUCLEONS AND HYPERONS  \\ AS CHIRAL SOLITONS IN THE NJL
MODEL\thanks{Supported by the Deutsche Forschungsgemeinschaft (DFG)
under contract number Re--856/2-2.}
}
\author{
R.\ ALKOFER, H.\ WEIGEL\thanks{Supported by a Habilitanden scholarship
of the DFG.}, H.\ REINHARDT AND A.\ ABADA \\
{\it Institute for Theoretical Physics, T\"ubingen University}\\
{\it Auf der Morgenstelle 14, 72076 T\"ubingen, Germany}
}
%%==============================================================
\maketitle

\begin{abstract}
In the chirally symmetric Nambu--Jona-Lasinio model baryons are
described as chiral solitons of mesonic quark--antiquark bound states.
Hyperons are investigated in two complementary pictures: the bound state
approach of Callan and Klebanov and the collective approach of Yabu and
Ando. The latter is used to compute the strange vector form factors
of the nucleon providing estimates for the strange electric mean
square radius and the strange magnetic moment of the nucleon.
% appr. 100 words, limit is 200
\end{abstract}

For a large number of colors ($N_C)$ QCD reduces to an effective
theory of weakly interacting mesons. Witten conjectured that within this
effective theory baryons emerge as soliton solutions. Although Witten's
conjecture has never been proven the soliton picture of baryons has turned
out quite successful in recent years.

The numerous attempts to derive the effective meson theory from QCD
indicate that at low energies the effective meson theory is almost
entirely determined by chiral symmetry. This suggests to
study simpler chirally invariant models of the quark flavor dynamics.
The prototype is the Nambu--Jona--Lasinio (NJL) model. It can be
rewritten in mesonic degrees of freedom (bosonization). The predictions
from the resulting effective meson theory are in satisfactory agreement
with the low-energy data for pseudoscalar mesons. Moreover, the
bosonized NJL model contains soliton solutions \cite{1,2}. The
studies of the soliton transparently explain how baryons emerge as
solitons in effective meson theories starting from an underlying quark
theory simultaneously confirming Witten's conjecture. The reason being
that the baryon number is carried by the polarized vacuum for the
self--consistent soliton when not only the pseudoscalar but also the
(axial)vector mesons are incorporated.

Aiming at the description of hyperons within the soliton
picture the flavor symmetry breaking has been treated in two
conceptionally different ways: In the bound state approach of Callan and
Klebanov (CK) and in the collective approach of Yabu and Ando (YA),
{\it cf.} table 1.
\begin{table}
\tcaption{ Comparison
of the collective and the bound state approach.}
{}~\newline
\vspace{-0.7cm}
\centerline{
\begin{tabular}{|c|c|}
\hline
Collective approach (YA) & Bound state approach (CK) \\
\hline
Symmetry breaking small & Symmetry breaking large \\
(light strange quark) & (heavy strange quark)\\
$\downarrow$ & $\downarrow$ \\
Strange components as & Restoring force for \\
{\bf collective coordinates} & {\bf strange fluctuations} \\
(analogous to zero modes)    & (``harmonic'' potential) \\
$\downarrow$ & $\downarrow$ \\
Collective Hamiltonian & Bound State Energy and \\
in flavor SU(3) including & Wave Function \\
symmetry breaking & $\downarrow$ \\
$\downarrow$ & Collective quantization of \\
Exact diagonalization & spin and isospin \\
$\downarrow$ & $\downarrow$ \\
{\it HYPERONS} &{\it HYPERONS} \\
\hline
\end{tabular}
}
\end{table}
They not only predict the spectrum of the low--lying
baryons but also yield baryon wave--functions in a certain
configuration space. This in turn permits studying the effects of
strangeness in the nucleon by computing the pertinent matrix elements.
This is very interesting because there are many experimental
indications for significant effects of strangeness in the nucleon.
Especially experiments measuring parity violating asymmetries in
scattering processes of polarized electrons on nuclei provide
access to hadronic ``observables" like
$\langle N|{\bar s}\gamma_\mu s|N\rangle$, which {\it e.g.} enter
the matrix elements of the neutral current between nucleon states.

For these studies we will consider the NJL model with scalar ($S$)
and pseudoscalar ($P$) fields only. After integrating out the quark
fields in favor of these mesons the action in Euclidean space reads
($M=S+iP$)
\begin{eqnarray}
A_{\rm NJL}=N_C {\rm Tr}_\Lambda{\rm log}
\left[i\partial\!\!\!/ _E
+\frac{1}{2}\left(M^{\dag}+M\right)
+\frac{1}{2}\gamma_5\left(M^{\dag}-M\right)\right]
+\frac{1}{4G}{\rm Tr}\left[{\hat m}^0\left(M^{\dag}+M\right)\right] ,
%\quad iD\!\!\!\!/ _E=i\partial\!\!\!/ _E
%+\frac{1}{2}\left(M^{\dag}+M\right)
%+\frac{1}{2}\gamma_5\left(M^{\dag}-M\right)
\label{eq1}
\end{eqnarray}
where the trace includes Euclidean space--time, flavor and spin
degrees of freedom. The proper--time regularization is indicated
by the $O(4)$ invariant cut--off $\Lambda$ and $G$ denotes the
dimensionful coupling constant of the NJL Model. The flavor
symmetry breaking occurs in two instances: explicitly in the current
quark mass matrix\footnote{We assume isospin symmetry
$m^0_u=m^0_d=m^0$.} ${\hat m}^0={\rm diag}(m^0,m^0,m^0_s)$ and
(as a consequence of the gap--equations) in the vacuum expectation
values of the meson fields $\langle M\rangle={\rm diag}(m,m,m_s)$.
The quantities $m$ and $m_s$ are referred to as the up and strange
constituent quark masses, respectively. The parameters in the
non--strange sector are fitted to the pion mass ($m_\pi=135$MeV)
and decay constant ($f_\pi=93$MeV). The remaining undetermined
parameter can be expressed in terms of $m$. In the strange sector
the kaon mass ($m_k=495$MeV) is employed to compute $m_s^0$. Then
the kaon decay constant ($f_k$) is left as a prediction. For commonly
accepted values of $m=350\ldots450$MeV the experimental value
($f_k\approx113$MeV) is underestimated by about 10 to 15\%.

In order to construct the static soliton solution to the action
(\ref{eq1}) the scalar fields are constrained to their vacuum expectation
values and for the pseudoscalar fields the hedgehog {\it ansatz} is
adopted
\begin{eqnarray}
M_0(\mbox{\boldmath $r$})=\pmatrix{
m\ {\rm exp}\left(i\mbox{\boldmath $\tau$}\cdot
\hat{\mbox{\boldmath $r$}}\Theta(r)\right) &
{| \atop |}&\hspace{-10pt}{\mbox{\small $0$}
\atop \mbox{\small $0$}}\cr
-------\ -\hspace{-8pt}&-&\hspace{-10pt}-----\cr
0\qquad 0&|&\hspace{-8pt} m_s\cr }.
\label{eq2}
\end{eqnarray}
The radial function $\Theta(r)$ minimizes the static energy associated
with (\ref{eq2}).

In the CK approach the $3\times3$ matrix $M$  not only contains
the hedgehog (\ref{eq2}) in the isospin subgroup but also a time
dependent kaon field $K(\mbox{\boldmath $r$},t)$. Futhermore collective
coordinates $R_I(t)$ are introduced for the iso--rotations. The latter
correspond to large amplitude fluctuations. Expanding the action
(\ref{eq1}) up to quadratic order in $K$ yields the associated
Bethe--Salpeter equation in the soliton background. This equation
contains a bound solution, {\it i.e.} its energy $\omega$ lies between
zero and $m_k$, which carries strangeness $S=-1$. Hence each occupation
of this bound state increases the energy by $|\omega|$ while $S$
decreases by one unit. Canonical quantization of $R_I$ finally
removes the degeneracy of baryons with identical spin and/or isospin
like $\Sigma$ and $\Lambda$. In the YA approach the kaon fields
are treated as large amplitude fluctuations as well, {\it i.e.}
$M(\mbox{\boldmath $r$},t)\sim
R_3(t)M_0(\mbox{\boldmath $r$})R_3(t)^{\dag}$, with $R_3(t)\in$SU(3).
These coordinates are canonically quantized yielding an Hamiltonian
which may be diagonalized exactly despite of the fact that symmetry
breaking parts occur. The predictions for the mass differences
are displayed in table 2. The CK approach seems to give
a somewhat better agreement with the experimental data.
\begin{table}
\tcaption{The mass differences of the low-lying
$\frac{1}{2}^+$
and $\frac{3}{2}^+$ baryons with respect to the nucleon. We compare
the predictions of two approaches to the NJL model with the experimental
data.  All numbers are in MeV. (Taken from ref.\ \protect{\cite{1}}.)
$m$ is chosen to reproduce $M_\Delta-M_N$.}
{}~\newline
\vspace{-0.8cm}
\centerline{
\begin{tabular}{c|c c|c||c}
{}~~~~~~~~~~& ~~~~~CK~~~~~ & ~~~~~YA~~~~~ & ~~~~~Expt.~~~~~
&~~YA~$f_k^{\rm corr}$~~ \\
$m$       & 430 &  407  & --- & 390 \\
\hline
$\Lambda$ & 132 &  105 & 177  & 175 \\
$\Sigma$ & 234  & 148 & 254  & 248 \\
$\Xi$ & 341 &  236 & 379  & 396 \\
$\Delta$ & 293  & 293 & 293  & 291 \\
$\Sigma^*$ & 374 & 387 & 446  & 449 \\
$\Xi^*$ & 481  & 482 & 591   & 608 \\
$\Omega$ & 613  & 576 & 733  & 765 \\
\end{tabular}}
\vspace{-0.5cm}
\end{table}
In both approaches the mass differences are slightly underestimated.
This is inherited from the meson sector where $f_k$ is predicted too
small. Adjusting for this shortcoming in the baryon sector as well
yields excellent agreement with the experiment\cite{1} (last column
in table 2).

In the next step the Noether currents associated with the action
(\ref{eq1}) are constructed. We are interested in the matrix elements
of the strange vector current between nucleon states
\begin{eqnarray}
\langle N|{\bar s}\gamma_\mu s|N\rangle=
{\bar u}(\mbox{\boldmath $p$}^\prime)\left[\gamma_\mu F_3(Q^2)
+\frac{\sigma_{\mu\nu}Q^\nu}{2M_N}\tilde F_3(Q^2)\right]
u(\mbox{\boldmath $p$})\ , \quad Q_\mu=p^\prime_\mu-p_\mu\ .
\label{eq3}
\end{eqnarray}
In the YA approach we have computed\cite{3} the
strange magnetic moment $\mu_s=\tilde F_3(0)$ and the mean squared
radius $\langle r_S^2\rangle$ of the electric form factor
$G_E=F_3+(Q^2/4M^2)\tilde F_3$. We predict
\begin{eqnarray}
-0.05\le\mu_s\le0.25 \qquad {\rm and} \qquad
-0.25{\rm fm}^2\le\langle r_S^2\rangle\le-0.15{\rm fm}^2.
\label{eq4}
\end{eqnarray}
The uncertainties stem from the fact that the NJL model result for
the isovector magnetic moment of the nucleon is too small and
pertinent adjustments (which are outside the self--consistent solution)
are necessary.

We have seen that the NJL model provides illuminating inside into
the description of baryons as solitons when starting from a
microscopic theory of the quark flavor dynamics.

\end{document}